# *Ab Initio* Study of Different Acid Molecules Interacting with $H_2O$

Aleksey A. Zakharenko, S. Karthikyan, K.S. Kim

**Address**: Center for Superfunctional Materials, Department of Chemistry, Pohang University of Science and Technology, Pohang 790-784, Korea.

**E-mail**: kim@postech.ac.kr

## Abstract

Using the Gaussian-03 for *ab initio* calculations, we have studied interaction of different acid molecules with a single water molecule. The molecular and supermolecular optimized structures were found with the Becke-3-Lee-Yang-Parr (B3LYP-hybrid potential) calculations of density-functional theory (DFT) methods as well as the Møller-Plesset second-order perturbation theory, using the basis set of Aug-cc-*p*VDZ quality and the CRENBL ECP effective core potential for molecules containing heavy iodine atom. Possible isomers of studied acids and supermolecules, consisting of acid molecules coupled with a single water molecule, are shown. Energies, zero-point energies (ZPEs), thermal enthalpies and free energies, as well as the corresponding binding energies for the theoretical methods were calculated. It was found that optimized structures of supermolecular isomers with lowest energies corresponding to the global minimum on the potential energy surfaces can be different for both theories. The simplest structure acids $H_2S$ and $H_2Se$, forming acid-water supermolecules, can give clear evidence of disagreement of the two theoretical methods concerning optimization of lowest energy structures, because the B3LYP-DFT method gives the lowest-energy structure for the first supermolecular isomer, but the MP2 method for the second possible isomer. A dramatic difference between potential energy surfaces for both theories applying to the optimized structure finding of the $H_2SO_3$-$H_2O$ supermolecular isomers was found, because MP2 supermolecular geometries cannot exist for the corresponding B3LYP-DFT ones, for which the frequency characteristics of the supermolecular isomers were also calculated. In general, the binding energies and ZPE ones for the MP2 method are 10-15% larger than those for the B3LYP-DFT method. However, the thermal free energies for the MP2 method can be significantly smaller than those for the B3LYP-DFT method.

## 1. Introduction

One of today recurrent topics of physical chemistry and chemical physics is the theoretical investigation of gas-phase clusters [1-4], including hydrogen-bonded complexes. The interaction of an acid molecule with water can lead to many different structures that could depend on chosen level of theory. It is well-known that the hydration process has important implications in the context of atmospheric chemistry and aerosol, because there occur reaction mechanisms of introduction of substantial amount of gas phase chlorine and bromine compounds into the marine troposphere [5-6]. The development of supersonic jet nozzles allowed extensive studying of different molecular clusters in complicated experiments, for instance, see in Refs. [7-11]. On the other hand, *ab initio* theoretical studies are widely used in addition to experimental investigations. Thus, a large set of the structural and thermochemical data were obtained in this theoretical study with quantum-mechanical methods.

In Ref. [12] was mentioned that the available structural, spectroscopic, and thermochemical data are still limited for the majority of hydrated halides. That is also true for chlorine-, bromine-, and iodine-containing acids, the recent theoretical and experimental investigations of which can be found in refs. [13-29]. Hypochlorous (HClO), hypobromous (HBrO), and hypo-iodous (HIO) acids are, probably, the simple examples and represent weak acids. Chlorous ($HClO_2$), bromous ($HBrO_2$), and iodous ($HIO_2$) acids of relatively weak-acid family are also well-known. Chloric acid with the chemical formula $HClO_3$ is known as a strong acid with $pK_a$ ~ - 1 and oxidizing agent. Bromic acid with the chemical formula $HBrO_3$ is a key reagent in the well-known Belousov-Zhabotinsky oscillating reaction and has about 62% bromine, 1% hydrogen, and 37% oxygen. Iodic acid with the chemical formula $HIO_3$ can be obtained as a white solid and is an insoluble compound, unlike chloric acid or bromic acid. Perchloric acid ($HClO_4$) is an oxoacid of chlorine and is a colorless liquid soluble in water. Perchloric acid is a strong super-acid completely dissociating in an aqueous solution comparable in strength to sulfuric acid ($H_2SO_4$) or nitric acid (). Perbromic acid ($HBrO_4$) is a strong acid and strongly oxidizing, and is the least stable of the halogen (VII) acids. Per-iodic acid ($HIO_4$) has heavy iodine atom and is widely used in organic chemistry for structural analysis.

On the other hand, there is much work on experimental and theoretical investigations of strong nitric super-acid ($HNO_3$) that is a highly corrosive and toxic acid that can cause severe burns. Example interesting theoretical and experimental studies of nitric acid can be found in Refs. [30-36] and [37-42], respectively. Also, there exist much theoretical (in general, density-functional theory or DFT) and experimental work on studying the strong sulfur-containing super-acid, $H_2SO_4$, see Refs. [43-46] as well as Refs. [47-50] of DFT-study. It is thought that due to experimental difficulties in investigations of the Sulfurous acid $H_2SO_3$, less attention is paid to the acid study [51-52], and it is little known about the so-called Sulfonic acid, which has the same chemical formula $H_2SO_3$. The sulfur-containing acid with the simplest chemical formula $H_2S$ (Sulfane or Hydrogen sulfide) is a covalent hydride structurally related to water ($H_2O$) since oxygen and sulfur occur in the same periodic table group. Hydrogen sulfide is weakly acidic, and references on the acid studies can be found in [53-56]. Hydrogen selenide ($H_2Se$) is the simplest hydride of selenium studied in Ref. [57], a colorless, flammable gas under standard conditions, and soluble in water. The properties of $H_2S$ and $H_2Se$ are similar. The well-known phosphorous-containing acids, phosphoric acid ($H_3PO_4$) and phosphorous acid ($H_3PO_3$), were studied in Refs. [58-62]. Phosphoric acid ($H_3PO_4$) also known as orthophosphoric acid or phosphoric (V) acid, is an inorganic acid and very commonly used as an aqueous solution. Phosphoric acid is also used as the electrolyte in phosphoric-acid fuel cells. Phosphorous acid [63-64], also called phosphonic acid, is one of the well-known and commonly used oxoacids of phosphorus. The other well-known acids are boric acid ($H_3BO_3$ or $B(OH)_3$) also called boracic acid or orthoboric acid**,** and boron oxide hydroxide (metaboric acid) with the chemical formula $HBO_2$. There is much work on the $H_3BO_3$ and $HBO_2$ acids used in medicine, for instance, see Refs. [65-70]. A highly valuable precursor to many chemical compounds ranging from polymers to pharmaceuticals, hydrogen cyanide is a chemical compound with the following chemical formula HCN. A solution of hydrogen cyanide in water is called hydrocyanic acid. Hydrogen cyanide is a colorless, very poisonous, and highly volatile liquid that boils slightly above room temperature at 26 °C (78.8 °F). Recent studies on the weakly acidic hydrogen cyanide can be found in Refs. [71-73]. It is noted that HCN as a simplest nitrile system can act as both proton donor and proton acceptor systems in cluster formation that can be the case for water and alcohol molecules. The interaction of HCN with water can leads to two different structural motifs.

## 2. *Ab initio* calculations

All the calculations were performed using the GAUSSIAN-03 program [74], see also the famous books [75-77]. The minimum-energy molecular structures of all supermolecular isomers were completely optimized by using density-functional theory (DFT [78]) calculations employing Becke's three-parameter exchange potential [79-80] and the Lee-Yang-Parr correlation functional [81] (B3LYP) as well as the Møller-Plesset [82] second-order perturbation theory (MP2) calculations employing Dunning's augmented basis set [83-84] of the Aug-cc-*p*VDZ quality. In a supermolecule consisting of two molecules (for instance, in the water dimer) basis functions from one molecule can help compensate for the basis set incompleteness on the other molecule, and vice versa. This effect is known as basis set superposition error (BSSE) which will be zero in the case of a complete basis set. An approximate way of assessing the BSSE is the counterpoise (CP) correction method [85-86], in which the BSSE can be estimated as the difference between monomer energies with the regular basis and the energies calculated with the full set of basis functions for the whole supermolecules studied in this paper. In addition to the binding energies ($\Delta E$), zero-point energies ($\Delta ZPE$), the enthalpies ($\Delta H$), and the Gibbs free energies ($\Delta G$) at 298K and 1 atm are also reported for both the B3LYP and MP2 theoretical methods.

Average relativistic effective potential (AREP) and spin-orbit (SO) operators for the chemical elements of the second transition row have been published in Ref. [87], in which particular attention was focused on the portioning of the core and valence space, as well as Gaussian basis sets with contraction coefficients for the lowest energy state of each atom were introduced. Discussion of the details and complete review were given in Ref. [88]. More general reviews on the subject of effective core potentials can be also found in the literature [89-90]. The effective core potential of Ref. [87] called CRENBL ECP was used in this report to calculate structural and energetic properties of (super)-molecules containing the heavy iodine atom possessing the atomic number 53 in the periodic table of chemical elements of Mendeleev.

## 3. Results and discussion

It was found with the B3LYP-DFT methods using the basis set of Aug-CC-*p*VDZ quality that three possible isomers for the chemical formula $H_2SO_3$ can exist. The most stable of them, the so-called Sulfurous acid ($H_2SO_3$) possesses the lowest energy on the potential energy surface, and its molecular structure is shown in Figure 1a with both hydrogen atoms rotated towards the third molecular oxygen, which is not coupled with a hydrogen atom. The second, less stable $H_2SO_3$ isomer has its ground state energy by about 1 kcal/mol smaller and opposite-directed hydrogen atoms (see the Table and Figure 1). The so-called Sulfonic acid being the third possible $H_2SO_3$ isomer is the most "unstable" with its local minimum energy value of by about  Hartrees that is ~ 1 kcal/mol smaller. However, a Sulfonic acid single molecule can have the strongest coupling with a single water molecule ($H_2O$) with a 10% larger binding energy $E_B$ compared with the other two isomers. Indeed, it is possible to suggest a 10% difference in energy can be treated as non-significant, and hence, one deals here with an isoenergetic case like studied in Refs. [Kim, others?]. Using the Moeller-Plesset second-order perturbation theory (MP2) [], obtained binding energies are always larger than corresponding those calculated at the hybrid B3LYP-DFT level of theory. Moreover, some optimized structures shown in Figure 1 cannot be found with the MP2-method, for instance, the structure shown in Table 1 with the largest binding energy Eb for the most stable $H_2SO_3$ isomer. That can be explained by the way that the potential energy surfaces (PESs) for both commonly-used methods can give even qualitatively different results concerning structure optimization. Also, for Sulfonic acid it was found that the water oxygen can interact with the $H_2SO_3$ molecular hydrogen coupled with sulfur atom. That is natural because an $H_2O$-molecule can readily interact with an $H_2S$-molecule [], if a sulfur atom is placed instead of an oxygen atom in a water dimer. However, in the case of O=HS-interaction for the $H_2SO_3$-$H_2O$ complex system the twice binding energy can be obtained compared with the relatively more simple system of O-HS-interaction: an $H_2S$-$H_2O$-system.

Calculations were done with the B3LYP-DFT method, using the basis set of Aug-CC-*p*VDZ quality. Energy of water monomer is $E_w$(B3LYP-DFT/Aug-CC-*p*VDZ) = – 76.4446427 atomic units (Hartrees). It is noted that the energy difference between the parallel-like and anti-parallel-like hydrogen orientations for $H_3PO_4$ is as high as the following value: $\Delta E_M$($H_3PO_4$, B3LYP-DFT/Aug-CC-*p*VDZ) = $E_M$(anti-parallel) – $E_M$(parallel) ~ 0.90 kcal/mol. The same value for Phosphonic acid $H_3PO_3$ is as follows: $\Delta E_M$($H_3PO_3$, B3LYP-DFT/Aug-CC-*p*VDZ) = $E_M$(anti-parallel) – $E_M$(parallel) = ~ 1.79 kcal/mol. This difference between the parallel-like and anti-parallel-like hydrogen orientations for Sulfuric acid $H_2SO_4$ two possible hydrogen orientations is as high as the following value: $\Delta E_M$($H_2SO_4$, B3LYP-DFT/Aug-CC-*p*VDZ) = $E_M$(anti-parallel) – $E_M$(parallel) ~ – 1.13 kcal/mol. The same value for Sulfurous acid $H_2SO_3$ is as follows: $\Delta E_M$($H_2SO_3$, B3LYP-DFT/Aug-CC-*p*VDZ) = $E_M$(anti-parallel) – $E_M$(parallel) ~ 0.93 kcal/mol. Using the MP2 calculations, these values for $H_3PO_4$ are as follows: $\Delta E_M$($H_3PO_4$, MP2/Aug-CC-*p*VDZ) ~ 0.98 kcal/mol; the value for Phosphonic acid $H_3PO_3$ is as follows: $\Delta E_M$($H_3PO_3$, MP2/Aug-CC-*p*VDZ) ~ 1.88 kcal/mol; for Sulfuric acid $H_2SO_4$ one can calculate $\Delta E_M$($H_2SO_4$, MP2/Aug-CC-*p*VDZ) ~ – 1.17 kcal/mol; for Sulfurous acid $H_2SO_3$ is as follows: $\Delta E_M$($H_2SO_3$, MP2/Aug-CC-*p*VDZ) ~ 0.81 kcal/mol.

The interaction energies $E_B$ of different molecules with a single water molecular, where $E_T$ represents the total energy of a sample molecule plus a single water molecule, $E_C$ is the corrected energy with the basis set superposition error (BSSE) correction, $E_{BSSE}$. Using the values of $E_{BSSE}$ and

$E_{BC} = E_C - E_M - E_w$,

the binding energy

$E_{B\Delta} = (E_B + E_{BC} \pm E_{BSSE})/2$

BSSE from the book

**4. Conclusions**

In this report, we have investigated the structural and energetic characteristics of the interaction of single acid molecules with a single water molecule. Both commonly-used theoretical methods, the Becke's three-parameter exchange potential with the Lee-Yang-Parr correlation functional and the second-order Møller-Plesset perturbation theory, have shown good agreement for optimized structures' finding and binding energies' calculations, using the same Dunning's basis set of Aug-cc-*p*VDZ quality and the CRENBL ECP effective core potential for molecules containing heavy iodine atom. However, some MP2 optimized structures of the acid-water supermolecules cannot be found for the corresponding B3LYP-DFT calculations that can be explained by the fact of uniqueness of each potential energy surface for the theoretical methods. That is even dramatic for the Sulfurous acid-water ($H_2SO_3$-$H_2O$) supermolecular isomers, because the corresponding MP2 optimized structures cannot be found for the B3LYP-DFT ones with the smallest and highest (most stable structure) binding energies.

**Acknowledgements**

It is a pleasure to acknowledge the Pohang University of Science and Technology for support from a POSTECH grant.

**Table 1.** The energies of molecules and supermolecules as well as binding energies. The lowest energies and lowest binding energies are shown bold, and the MP2 calculations are shown italic compared with the B3LYP-DFT calculations. The MP2 calculations are given above the corresponding B3LYP-DFT energies. Note that some MP2 structures for the B3LYP ones were not found, and hence, their energies were not introduced in the Table. Here the energies (E), zero point energies (ZPE), thermal enthalpies (H), and thermal free energies (G) are given in the Hartree energy units, and the corresponding binding energies ΔE, ΔZPE, ΔH, and ΔG are given in kcal/mol. The CRENBL ECP effective core potential was used to calculate energies of supermolecules containing the heavy iodine atom. In the first column of the table, the labels in the parentheses for the (super)-molecular isomers corresponds to those for the (super)-molecular structures shown in Figures 1 and 2.

| Name | E, Hartrees | ZPE | H | G | ΔE, kcal/mol | ΔZPE | ΔH | ΔG |
|---|---|---|---|---|---|---|---|---|
| *$H_2O$(MP2)* | *-76.26091* | *-76.23958* | *-76.23580* | *-76.25723* | - | - | - | - |
| $H_2O$ | -76.44464 | -76.42341 | -76.41963 | -76.44106 | - | - | - | - |
| *$H_2S0_4$(MP2)* | *-699.06475* | *-699.02675* | *-699.02045* | *-699.05525* | - | - | - | - |
| $H_2S0_4$(trans) | **-700.28055** | **-700.24300** | **-700.23659** | **-700.27159** | - | - | - | - |
| *$H_2S0_4$-$H_2O$(MP2)* | *-775.34588* | *-775.28290* | *-775.27385* | *-775.31551* | *-12.69* | ***-10.40*** | *-11.05* | ***-1.903*** |
| $H_2S0_4$-$H_2O$(1U11t) | -776.74293 | -776.68058 | -776.67145 | -776.71327 | **-11.13** | **-8.89** | **-9.56** | **-0.385** |
| *$H_2S0_4$-$H_2O$(MP2)* | *-775.34595* | *-775.28288* | *-775.27387* | *-775.31548* | ***-12.73*** | *-10.39* | ***-11.06*** | *-1.879* |
| $H_2S0_4$-$H_2O$(1U'11t) | -776.74287 | -776.68052 | -776.67139 | -776.71327 | -11.09 | -8.85 | -9.51 | -0.384 |
| | | | | | | | | |
| *$H_2S0_4$(MP2)* | *-699.06289* | *-699.02522* | *-699.01867* | *-699.05429* | - | - | - | - |
| $H_2S0_4$(cis) | -700.27875 | -700.24148 | -700.23486 | -700.27064 | - | - | - | - |
| *$H_2S0_4$-$H_2O$(MP2)* | *-775.34407* | *-775.28120* | *-775.27206* | *-775.31432* | ***-11.55*** | ***-9.34*** | ***-9.92*** | ***-1.15*** |
| $H_2S0_4$-$H_2O$(1U11c) | -776.74119 | -776.67894 | -776.66975 | -776.71206 | **-10.04** | **-7.86** | **-8.49** | **0.37** |
| *$H_2S0_4$-$H_2O$(MP2)* | *-775.34380* | *-775.28045* | *-775.27151* | *-775.31263* | *-11.38* | *-8.87* | *-9.58* | *-0.09* |
| $H_2S0_4$-$H_2O$(1U12c) | -776.73931 | -776.67662 | -776.66750 | -776.70970 | -8.86 | -6.40 | -7.08 | 1.85 |
| *$H_2S0_4$-$H_2O$(MP2)* | *-775.33002* | *-775.26908* | *-775.25847* | *-775.30624* | *-2.74* | *-1.73* | *-1.40* | *3.92* |
| $H_2S0_4$-$H_2O$(1U10c) | -776.72768 | -776.66750 | -776.65661 | -776.70670 | -1.56 | -0.69 | -0.24 | 3.74 |
| | | | | | | | | |
| *$H_2SO_3$(MP2)* | *-624.01878* | *-623.98708* | *-623.98107* | *-624.01499* | - | - | - | - |
| $H_2SO_3$(p) | **-625.07765** | **-625.04605** | **-625.04003** | **-625.07395** | - | - | - | - |
| *$H_2SO_3$-$H_2O$(MP2)* | *-700.29837* | *-700.24089* | *-700.23255* | *-700.27197* | ***-11.72*** | ***-8.93*** | ***-9.84*** | ***0.16*** |
| $H_2SO_3$-$H_2O$(1U12sp) | -701.53721 | -701.48016 | -701.47170 | -701.51191 | -9.36 | -6.71 | -7.55 | 1.95 |
| *$H_2SO_3$-$H_2O$(MP2)* | *-700.29665* | *-700.23982* | *-700.23114* | *-700.27185* | *-10.64* | *-8.26* | *-8.95* | *0.23* |
| $H_2SO_3$-$H_2O$(1U'11sp) | -701.53700 | -701.48042 | -701.47177 | -701.51228 | -9.23 | -6.87 | -7.60 | 1.71 |
| *$H_2SO_3$-$H_2O$(MP2)* | - | - | - | - | - | - | - | - |
| $H_2SO_3$-$H_2O$(1U01sp) | -701.52625 | -701.47145 | -701.46138 | -701.50787 | -2.48 | -1.24 | -1.08 | 4.49 |
| *$H_2SO_3$-$H_2O$(MP2)* | - | - | - | - | - | - | - | - |
| $H_2SO_3$-$H_2O$(1U00sp) | -701.52244 | -701.46871 | -701.45785 | -701.50718 | -0.09 | 0.48 | 1.14 | 4.92 |
| *$H_2SO_3$-$H_2O$(MP2)* | - | - | - | - | - | - | - | - |
| $H_2SO_3$-$H_2O$(1U11sp) | -701.53796 | -701.48116 | -701.47264 | -701.51291 | **-9.83** | **-7.34** | **-8.15** | **1.32** |
| *$H_2SO_3$-$H_2O$(MP2)* | *-700.28656* | *-700.23131* | *-700.22154* | *-700.26485* | *-4.31* | *-2.92* | *-2.93* | *4.63* |
| $H_2SO_3$-$H_2O$(1U02sp) | -701.52574 | -701.47097 | -701.46092 | -701.50615 | -2.16 | -0.94 | -0.79 | 5.56 |
| *$H_2SO_3$-$H_2O$(MP2)* | *-700.28563* | *-700.23057* | *-700.22062* | *-700.26629* | *-3.73* | *-2.46* | *-2.35* | *3.73* |
| $H_2SO_3$-$H_2O$(1U'01sp) | - | - | - | - | - | - | - | - |
| | | | | | | | | |
| *$H_2SO_3$(MP2)* | *-624.01749* | *-623.98565* | *-623.97971* | *-624.01334* | - | - | - | - |
| $H_2SO_3$(ap) | -625.07617 | -625.04441 | -625.03846 | -625.07211 | - | - | - | - |
| *$H_2SO_3$-$H_2O$(MP2)* | *-700.29397* | *-700.23728* | *-700.22837* | *-700.26928* | *-8.96* | *-6.67* | *-7.22* | *1.85* |
| $H_2SO_3$-$H_2O$(1U02sa) | -701.53319 | -701.47688 | -701.46786 | -701.50915 | -6.84 | -4.65 | -5.14 | 3.68 |
| *$H_2SO_3$-$H_2O$(MP2)* | *-700.29657* | *-700.23945* | *-700.23090* | *-700.27104* | *-10.59* | *-8.03* | *-8.81* | *0.75* |
| $H_2SO_3$-$H_2O$(1U'11sa) | -701.53654 | -701.47969 | -701.47115 | -701.51126 | -8.94 | -6.41 | -7.21 | 2.36 |
| *$H_2SO_3$-$H_2O$(MP2)* | *-700.29389* | *-700.23716* | *-700.22826* | *-700.26922* | *-8.91* | *-6.59* | *-7.15* | *1.89* |
| $H_2SO_3$-$H_2O$(1U'02sa) | -701.53316 | -701.47681 | -701.46779 | -701.50911 | -6.82 | -4.61 | -5.10 | 3.71 |
| *$H_2SO_3$-$H_2O$(MP2)* | *-700.29709* | *-700.23983* | *-700.23135* | *-700.27133* | ***-10.92*** | ***-8.27*** | ***-9.09*** | ***0.56*** |
| $H_2SO_3$-$H_2O$(1U11sa) | -701.53692 | -701.47995 | -701.47146 | -701.51146 | **-9.17** | **-6.58** | **-7.41** | **2.23** |
| | | | | | | | | |
| *$H_2SO_3$(MP2)* | *-623.98171* | *-623.94863* | *-623.94338* | *-623.97576* | - | - | - | - |
| $H_2SO_3$(s) | **-625.03453** | **-625.00216** | **-624.99684** | **-625.02935** | - | - | - | - |
| *$H_2SO_3$-$H_2O$(MP2)* | *-700.26253* | *-700.20423* | *-700.19638* | *-700.23548* | ***-12.50*** | ***-10.06*** | ***-10.80*** | ***-1.56*** |
| $H_2SO_3$-$H_2O$(1U11su) | -701.49683 | -701.43945 | -701.43156 | -701.47073 | **-11.08** | **-8.70** | **-9.46** | **-0.20** |
| *$H_2SO_3$-$H_2O$(MP2)* | *-700.25228* | *-700.19628* | *-700.18671* | *-700.23177* | *-6.06* | *-5.07* | *-4.73* | *0.77* |
| $H_2SO_3$-$H_2O$(1U00su) | -701.48756 | -701.43234 | -701.42280 | -701.46718 | -5.26 | -4.25 | -3.97 | 2.03 |
| | | | | | | | | |
| *$H_3PO_4$(MP2)* | *-643.01739* | *-642.96931* | *-642.96206* | *-642.99877* | - | - | - | - |
| $H_3PO_4$(p) | **-644.21347** | **-644.16560** | **-644.15825** | **-644.19526** | - | - | - | - |
| *$H_3PO_4$-$H_2O$(MP2)* | *-719.30076* | *-719.22718* | *-719.21745* | *-719.26078* | ***-14.09*** | ***-11.48*** | ***-12.30*** | ***-2.99*** |
| $H_3PO_4$-$H_2O$(1U11pp) | -720.67652 | -720.60332 | -720.59356 | -720.63672 | **-11.55** | **-8.97** | **-9.84** | -0.25 |
| *$H_3PO_4$-$H_2O$(MP2)* | *-719.30044* | *-719.22692* | *-719.21711* | *-719.26024* | *-13.89* | *-11.32* | *-12.08* | *-2.65* |
| $H_3PO_4$-$H_2O$(1U'11pp) | -720.67610 | -720.60310 | -720.59316 | -720.63684 | -11.29 | -8.84 | -9.59 | **-0.32** |
| | | | | | | | | |
| *$H_3PO_4$(MP2)* | *-643.01896* | *-642.97086* | *-642.96357* | *-643.00046* | - | - | - | - |
| $H_3PO_4$(ap) | -644.21204 | -644.16431 | -644.15692 | -644.19394 | - | - | - | - |
| *$H_3PO_4$-$H_2O$(MP2)* | *-719.29911* | *-719.22582* | *-719.21603* | *-719.25942* | *-13.06* | *-10.63* | *-11.41* | *-2.15* |
| $H_3PO_4$-$H_2O$(1U"11pa) | -720.67510 | -720.60219 | -720.59237 | -720.63570 | -10.66 | -8.27 | -9.09 | 0.39 |
| *$H_3PO_4$-$H_2O$(MP2)* | *-719.29511* | *-719.22198* | *-719.21188* | *-719.25570* | *-10.54* | *-8.22* | *-8.80* | *0.19* |
| $H_3PO_4$-$H_2O$(1U02pa) | -720.67045 | -720.59778 | -720.58758 | -720.63178 | -7.74 | -5.50 | -6.09 | 2.85 |
| *$H_3PO_4$-$H_2O$(MP2)* | *-719.29956* | *-719.22610* | *-719.21632* | *-719.25931* | *-13.34* | *-10.80* | *-11.59* | *-2.07* |
| $H_3PO_4$-$H_2O$(1U'11pa) | -720.67544 | -720.60237 | -720.59257 | -720.63559 | -10.88 | -8.38 | -9.22 | 0.46 |
| *$H_3PO_4$-$H_2O$(MP2)* | *-719.29995* | *-719.22636* | *-719.21665* | *-719.25952* | ***-13.58*** | ***-10.97*** | ***-11.80*** | ***-2.21*** |
| $H_3PO_4$-$H_2O$(1U11pa) | -720.67581 | -720.60270 | -720.59292 | -720.63598 | **-11.10** | **-8.59** | **-9.44** | **0.22** |
| | | | | | | | | |
| *$H_3PO_3$(MP2)* | *-567.90391* | *-567.86147* | *-567.85460* | *-567.89029* | - | - | - | - |
| $H_3PO_3$ | -568.94357 | -568.90125 | -568.89438 | -568.93000 | - | - | - | - |
| $H_3PO_3$(ts) | **-568.94758** | **-568.90518** | **-568.89847** | **-568.93371** | -2.52 | - | - | - |
| *$H_3PO_3$-$H_2O$(MP2)* | *-644.18446* | *-644.11649* | *-644.10713* | *-644.14904* | *-12.32* | *-9.69* | *-10.50* | ***-0.95*** |
| $H_3PO_3$-$H_2O$(1U'11pt) | -645.40428 | -645.33671 | -645.32729 | -645.36926 | -7.56 | **-5.10** | -5.76 | 3.46 |
| *$H_3PO_3$-$H_2O$(MP2)* | *-644.17595* | *-644.10945* | *-644.09892* | *-644.14595* | *-6.99* | *-5.27* | *-5.35* | *0.98* |
| $H_3PO_3$-$H_2O$(1U"01pt) | -645.39751 | -645.33138 | -645.32078 | -645.36788 | -3.32 | -1.75 | -1.68 | 4.33 |
| *$H_3PO_3$-$H_2O$(MP2)* | - | - | - | - | - | - | - | - |
| $H_3PO_3$-$H_2O$(1U'01pt) | -645.40214 | -645.33612 | -645.32567 | -645.37153 | -6.22 | -4.72 | -4.74 | **2.04** |
| *$H_3PO_3$-$H_2O$(MP2)* | *-644.18322* | *-644.11566* | *-644.10618* | *-644.14827* | *-11.55* | *-9.17* | *-9.91* | *-0.47* |
| $H_3PO_3$-$H_2O$(1U01pt) | -645.40248 | -645.33611 | -645.32587 | -645.37113 | -6.43 | -4.72 | -4.87 | 2.29 |
| *$H_3PO_3$-$H_2O$(MP2)* | *-644.18466* | *-644.11665* | *-644.10732* | *-644.14901* | ***-12.45*** | ***-9.79*** | ***-10.62*** | *-0.93* |
| $H_3PO_3$-$H_2O$(1U11pt) | -645.40431 | -645.33669 | -645.32733 | -645.36905 | **-7.58** | -5.08 | **-5.79** | 3.59 |
| | | | | | | | | |
| *$H_3PO_3$(MP2)* | *-567.90943* | *-567.86678* | *-567.86054* | *-567.89518* | - | - | - | - |
| $H_3PO_3$(p) | **-568.94195** | **-568.89975** | **-568.89348** | **-568.92811** | - | - | - | - |
| *$H_3PO_3$-$H_2O$(MP2)* | *-644.17778* | *-644.11144* | *-644.10152* | *-644.14518* | *-4.67* | *-3.19* | *-3.25* | *4.54* |
| $H_3PO_3$-$H_2O$(1U02p) | -645.39076 | -645.32528 | -645.31505 | -645.36061 | -2.61 | -1.33 | -1.22 | 5.37 |
| *$H_3PO_3$-$H_2O$(MP2)* | *-644.17744* | *-644.11119* | *-644.10116* | *-644.14617* | *-4.46* | *-3.03* | *-3.03* | *3.92* |
| $H_3PO_3$-$H_2O$(1U01p) | -645.39128 | -645.32568 | -645.31552 | -645.36118 | -2.94 | -1.58 | -1.51 | 5.02 |
| *$H_3PO_3$-$H_2O$(MP2)* | *-644.19121* | *-644.12295* | *-644.11437* | *-644.15495* | ***-13.10*** | ***-10.41*** | ***-11.32*** | ***-1.59*** |
| $H_3PO_3$-$H_2O$(1U11p) | -645.40516 | -645.33742 | -645.32887 | -645.36922 | **-11.65** | **-8.95** | **-9.89** | **-0.03** |
| *$H_3PO_3$-$H_2O$(MP2)* | *-644.19094* | *-644.12270* | *-644.11410* | *-644.15458* | *-12.93* | *-10.26* | *-11.15* | *-1.36* |
| $H_3PO_3$-$H_2O$(1U'11p) | -645.40488 | -645.33722 | -645.32862 | -645.36904 | -11.47 | -8.82 | -9.73 | 0.08 |
| | | | | | | | | |
| *$H_3PO_3$(MP2)* | *-567.90643* | *-567.86383* | *-567.85764* | *-567.89197* | - | - | - | - |
| $H_3PO_3$(ap) | -568.93909 | -568.89695 | -568.89073 | -568.92508 | - | - | - | - |
| *$H_3PO_3$-$H_2O$(MP2)* | *-644.18361* | *-644.11602* | *-644.10697* | *-644.14854* | ***-8.33*** | *-6.06* | ***-6.68*** | *2.43* |
| $H_3PO_3$-$H_2O$(1U11a) | -645.39728 | -645.33043 | -645.32129 | -645.36343 | **-6.71** | **-4.56** | **-5.14** | **3.61** |
| *$H_3PO_3$-$H_2O$(MP2)* | *-644.18361* | *-644.11603* | *-644.10697* | *-644.14857* | *-8.32* | ***-6.07*** | ***-6.68*** | ***2.41*** |
| $H_3PO_3$-$H_2O$(1U'11a) | -645.39721 | -645.33040 | -645.32121 | -645.36313 | -6.66 | -4.54 | -5.08 | 3.79 |
| | | | | | | | | |
| *$H_3BO_3$(MP2)lower* | *-251.91840* | *-251.86965* | *-251.86432* | *-251.89611* | - | - | - | - |
| $H_3BO_3$ | -252.53406 | -252.48561 | -252.48015 | -252.51215 | - | - | - | - |
| $H_3BO_3$(lower) | -252.54088 | -252.49203 | -252.48672 | -252.51845 | -4.28 | - | - | - |
| *$H_3BO_3$-$H_2O$(MP2)* | *-328.19370* | *-328.11998* | *-328.11168* | *-328.15081* | *-9.03* | *-6.75* | *-7.26* | *1.59* |
| $H_3BO_3$-$H_2O$(1U11b) | -328.99748 | -328.92382 | -328.91553 | -328.95462 | -7.50 | -5.26 | -5.76 | 3.07 |
| | | | | | | | | |
| *$HBO_2$(MP2)* | *-175.57830* | *-175.55828* | *-175.55395* | *-175.58157* | - | - | - | - |
| $HBO_2$ | -176.01562 | -175.99542 | -175.99110 | -176.01867 | - | - | - | - |
| *$HBO_2$-$H_2O$(MP2)* | *-251.85549* | *-251.81075* | *-251.80334* | *-251.84109* | *-10.21* | *-8.09* | *-8.53* | *-1.43* |
| $HBO_2$-$H_2O$(1U01b) | -252.47583 | -252.43113 | -252.42371 | -252.46147 | -9.77 | -7.72 | -8.14 | -1.09 |
| | | | | | | | | |
| *HCN(MP2)* | *-93.18366* | *-93.16806* | *-93.16453* | *-93.18749* | - | - | - | - |
| HCN | -93.43705 | -93.42089 | -93.41738 | -93.44027 | - | - | - | - |
| *HCN-$H_2O$(MP2)* | *-169.45159* | *-169.41244* | *-169.40543* | *-169.44052* | *-4.41* | *-3.01* | *-3.20* | *2.64* |
| HCN-$H_2O$(1U'00n) | -169.88750 | -169.84787 | -169.84090 | -169.87586 | -3.64 | -2.24 | -2.44 | 3.43 |
| *HCN-$H_2O$(MP2)* | *-169.45359* | *-169.41480* | *-169.40746* | *-169.44282* | ***-5.66*** | ***-4.50*** | ***-4.48*** | ***1.19*** |
| HCN-$H_2O$(1U00n) | -169.88980 | -169.85063 | -169.84336 | -169.87821 | **-5.09** | **-3.97** | **-3.98** | **1.97** |
| | | | | | | | | |
| *$HNO_3$(MP2)* | *-280.28489* | *-280.25848* | *-280.25399* | *-280.28425* | - | - | - | - |
| $HNO_3$ | -280.94027 | -280.91394 | -280.90948 | -280.93969 | - | - | - | - |
| *$HNO_3$-$H_2O$(MP2)* | *-356.55047* | *-356.50089* | *-356.49221* | *-356.53662* | *-2.94* | *-1.78* | *-1.52* | *3.05* |
| $HNO_3$-$H_2O$(1U10n) | -357.38804 | -357.33866 | -357.33002 | -357.37293 | -1.96 | -0.82 | -0.57 | 4.91 |
| *$HNO_3$-$H_2O$(MP2)* | *-356.56223* | *-356.51103* | *-356.50347* | *-356.54149* | ***-10.31*** | ***-8.15*** | ***-8.59*** | ***0.00*** |
| $HNO_3$-$H_2O$(1U11n) | -357.40016 | -357.34940 | -357.34187 | -357.37981 | **-9.57** | **-7.56** | **-8.01** | **0.59** |
| *$HNO_3$-$H_2O$(MP2)* | - | - | - | - | - | - | - | - |
| $HNO_3$-$H_2O$(1U'11n) | -357.38765 | -357.33871 | -357.32981 | -357.37513 | -1.72 | -0.85 | -0.44 | 3.53 |
| | | | | | | | | |
| *$H_2S$(MP2)* | *-398.85322* | *-398.83789* | *-398.83410* | *-398.85812* | - | - | - | - |
| $H_2S$ | -399.41512 | -399.40022 | -399.39642 | -399.41980 | - | - | - | - |
| *H2S-H2O(MP2)* | *-475.11923* | *-475.08050* | *-475.07283* | *-475.10991* | *-3.20* | *-1.90* | *-1.84* | ***3.41*** |
| H2S-H2O(s1U00) | -475.86394 | -475.82531 | -475.81800 | -475.85414 | **-2.62** | **-1.06** | **-1.22** | 4.22 |
| *H2S-H2O(MP2)* | *-475.11960* | *-475.08052* | *-475.07312* | *-475.10967* | ***-3.43*** | ***-1.92*** | ***-2.02*** | *3.56* |
| H2S-H2O(s1U'00) | -475.86340 | -475.82528 | -475.81757 | -475.85482 | -2.28 | -1.04 | -0.95 | **3.79** |
| | | | | | | | | |
| *$H_2Se$(MP2)* | *-2401.13051* | *-2401.11656* | *-2401.11275* | *-2401.13829* | - | - | - | - |
| $H_2Se$ | -2402.76114 | -2402.74781 | -2402.74400 | -2402.76891 | - | - | - | - |
| *$H_2Se$-$H_2O$(MP2)* | *-2477.39602* | *-2477.35885* | *-2477.35101* | *-2477.39000* | *-2.88* | *-1.70* | *-1.54* | *3.47* |
| $H_2Se$-$H_2O$(se1U00) | -2479.20959 | -2479.17253 | -2479.16521 | -2479.20265 | **-2.39** | **-0.82** | **-0.99** | 4.59 |
| *$H_2Se$-$H_2O$(MP2)* | *-2477.39711* | *-2477.35950* | *-2477.35202* | *-2477.39012* | ***-3.57*** | ***-2.11*** | ***-2.18*** | ***3.39*** |
| $H_2Se$-$H_2O$(se1U'00) | -2479.20855 | -2479.17237 | -2479.16439 | -2479.20404 | -1.74 | -0.72 | -0.47 | **3.73** |
| | | | | | | | | |
| *HClO(MP2)* | *-535.14531* | *-535.13502* | *-535.13114* | *-535.15814* | - | - | - | - |
| *HClO(MP2)* | *-535.24445* | *-535.23138* | *-535.22749* | *-535.25437* | *-62.21** | - | - | - |
| HClO | -535.90020 | -535.89076 | -535.88682 | -535.91393 | - | - | - | - |
| HClO(lower) | -535.99291 | -535.97986 | -535.97597 | -536.00285 | -58.18* | - | - | - |
| *HClO-H$_2$O(MP2)* | *-611.51052* | *-611.47432* | *-611.46641* | *-611.50439* | *-3.24* | *-2.11* | *-1.96* | *4.53* |
| HClO-H$_2$O(c1U00) | -612.44142 | -612.40527 | -612.39740 | -612.43541 | -2.43 | -1.26 | -1.13 | 5.33 |
| *HClO-H$_2$O(MP2)* | *-611.51815* | *-611.48040* | *-611.47330* | *-611.51093* | ***-8.03*** | ***-5.92*** | ***-6.29*** | ***0.43*** |
| HClO-H$_2$O(c1U01) | -612.44900 | -612.41150 | -612.40439 | -612.44175 | **-7.18** | **-5.16** | **-5.52** | **1.36** |
| *HClO-H$_2$O(MP2)* | *-611.51762* | *-611.48003* | *-611.47284* | *-611.51044* | *-7.70* | *-5.70* | *-6.00* | *0.73* |
| HClO-H$_2$O(c1U'01) | -612.44866 | -612.41129 | -612.40410 | -612.44155 | -6.97 | -5.03 | -5.33 | 1.48 |
| *HClO-H$_2$O(MP2)* | - | - | - | - | - | - | - | - |
| HClO-H$_2$O(c1U"01) | -612.44182 | -612.40519 | -612.39757 | -612.43642 | -2.68 | -1.20 | -1.24 | 4.70 |
| | | | | | | | | |
| *HClO$_2$(MP2)* | *-610.18816* | *-610.17178* | *-610.16702* | *-610.19764* | - | - | - | - |
| HClO$_2$ | -611.11090 | -611.09481 | -611.09004 | -611.12068 | - | - | - | - |
| *HClO$_2$-H$_2$O(MP2)* | *-686.46664* | *-686.42488* | *-686.41756* | *-686.45463* | ***-11.02*** | ***-8.49*** | ***-9.25*** | ***0.15*** |
| HClO$_2$-H$_2$O(c'1U11) | -687.57126 | -687.53009 | -687.52275 | -687.55988 | **-9.87** | **-7.45** | **-8.21** | **1.17** |
| *HClO$_2$-H$_2$O(MP2)* | *-686.45729* | *-686.41738* | *-686.40891* | *-686.44910* | *-5.16* | *-3.78* | *-3.82* | *3.62* |
| HClO$_2$-H$_2$O(c'1U01) | -687.56005 | -687.52075 | -687.51203 | -687.55340 | -2.83 | -1.59 | -1.48 | 5.24 |
| *HClO$_2$-H$_2$O(MP2)* | *-686.45759* | *-686.41714* | *-686.40887* | *-686.44959* | *-5.34* | *-3.63* | *-3.80* | *3.31* |
| HClO$_2$-H$_2$O(c'1U10) | -687.56239 | -687.52259 | -687.51415 | -687.55583 | -4.30 | -2.74 | -2.81 | 3.71 |
| *HClO$_2$-H$_2$O(MP2)* | *-686.45686* | *-686.41712* | *-686.40847* | *-686.44953* | *-4.89* | *-3.62* | *-3.55* | *3.35* |
| HClO$_2$-H$_2$O(c"1U10) | -687.56127 | -687.52192 | -687.51322 | -687.55510 | -3.60 | -2.32 | -2.23 | 4.17 |
| *HClO$_2$-H$_2$O(MP2)* | *-686.46643* | *-686.42462* | *-686.41730* | *-686.45435* | *-10.90* | *-8.32* | *-9.09* | *0.32* |
| HClO$_2$-H$_2$O(c"1U11) | -687.57116 | -687.52998 | -687.52262 | -687.55979 | -9.80 | -7.38 | -8.13 | 1.22 |
| | | | | | | | | |
| *HClO$_3$(MP2)* | *-685.16792* | *-685.14761* | *-685.14185* | *-685.17556* | - | - | - | - |
| HClO$_3$ | -686.25447 | -686.23468 | -686.22896 | -686.26246 | - | - | - | - |
| *HClO$_3$-H$_2$O(MP2)* | *-761.43748* | *-761.39420* | *-761.38432* | *-761.42923* | *-5.43* | *-4.40* | *-4.18* | *2.24* |
| HClO$_3$-H$_2$O(c1U20) | -762.70492 | -762.66217 | -762.65230 | -762.69701 | -3.64 | -2.56 | -2.33 | 4.09 |
| *HClO$_3$-H$_2$O(MP2)* | *-761.44438* | *-761.39861* | *-761.39033* | *-761.43047* | ***-9.76*** | ***-7.17*** | ***-7.96*** | ***1.46*** |
| HClO$_3$-H$_2$O(c1U11) | -762.71406 | -762.66919 | -762.66092 | -762.70097 | **-9.38** | **-6.96** | **-7.74** | **1.60** |
| *HClO$_3$-H$_2$O(MP2)* | *-761.43935* | *-761.39526* | *-761.38601* | *-761.42847* | *-6.60* | *-5.07* | *-5.25* | *2.71* |
| HClO$_3$-H$_2$O(c'1U'01) | -762.70444 | -762.66140 | -762.65175 | -762.69539 | -3.34 | -2.08 | -1.98 | 5.11 |
| *HClO$_3$-H$_2$O(MP2)* | - | - | - | - | - | - | - | - |
| HClO$_3$-H$_2$O(c'1U'10) | -762.70421 | -762.66145 | -762.65159 | -762.69686 | -3.19 | -2.10 | -1.88 | 4.18 |
| | | | | | | | | |
| *HClO$_4$(MP2)* | *-760.12789* | *-760.10182* | *-760.09586* | *-760.13022* | - | - | - | - |
| HClO$_4$(Im) | -761.37488 | -761.35035 | -761.34483 | -761.37832 | - | - | - | - |
| HClO$_4$(lower) | -761.37538 | -761.35051 | -761.34434 | -761.37920 | -0.31* | - | - | - |
| *HClO$_4$-H$_2$O(MP2)* | *-836.40671* | *-836.35574* | *-836.34688* | *-836.38876* | *-11.24* | *-9.00* | *-9.55* | *-0.82* |
| HClO$_4$-H$_2$O(1U'c11) | -837.83628 | -837.78690 | -837.77781 | -837.82028 | -10.20 | -8.14 | -8.68 | -0.01 |
| *HClO$_4$-H$_2$O(MP2)* | *-836.40759* | *-836.35641* | *-836.34763* | *-836.38932* | ***-11.79*** | ***-9.42*** | ***-10.02*** | ***-1.17*** |
| HClO$_4$-H$_2$O(1Uc11) | -837.83693 | -837.78749 | -837.77835 | -837.82164 | **-10.61** | **-8.51** | **-9.02** | **-0.86** |
| *HClO$_4$-H$_2$O(MP2)* | - | - | - | - | - | - | - | - |
| HClO$_4$-H$_2$O(1Uc10) | -837.82204 | -837.77456 | -837.76395 | -837.81232 | -1.26 | -0.40 | 0.02 | 4.98 |
| | | | | | | | | |
| *HBrO(MP2)* | *-2648.03692* | *-2648.02755* | *-2648.02363* | *-2648.05194* | - | - | - | - |
| *HBrO(MP2)lower* | *-2648.13700* | *-2648.12430* | *-2648.12037* | *-2648.14850* | *-62.80** | - | - | - |
| HBrO | -2649.84942 | -2649.84097 | -2649.83699 | -2649.86543 | - | - | - | - |
| HBrO(lower) | -2649.94485 | -2649.93216 | -2649.92822 | -2649.95636 | -59.89* | - | - | - |
| *HBrO-H$_2$O(MP2)* | *-2724.40604* | *-2724.37000* | *-2724.36219* | *-2724.40063* | *-5.10* | *-3.84* | *-3.78* | *3.20* |
| HBrO-H$_2$O(b1U00) | -2726.39582 | -2726.35974 | -2726.35201 | -2726.39024 | -3.97 | -2.61 | -2.61 | 4.51 |
| *HBrO-H$_2$O(MP2)* | *-2724.41033* | *-2724.37308* | *-2724.36584* | *-2724.40560* | ***-7.79*** | ***-5.77*** | ***-6.07*** | ***0.09*** |
| HBrO-H$_2$O(b1U01) | -2726.40017 | -2726.36306 | -2726.35586 | -2726.39456 | **-6.70** | **-4.70** | **-5.02** | 1.80 |
| *HBrO-H$_2$O(MP2)* | *-2724.40994* | *-2724.37277* | *-2724.36549* | *-2724.40447* | *-7.55* | *-5.58* | *-5.85* | *0.79* |
| HBrO-H$_2$O(b1U'01) | -2726.39995 | -2726.36299 | -2726.35569 | -2726.39461 | -6.56 | -4.66 | -4.92 | **1.77** |
| *HBrO-H$_2$O(MP2)* | *-2724.40489* | *-2724.36825* | *-2724.36073* | *-2724.40061* | *-4.38* | *-2.74* | *-2.87* | *3.22* |
| HBrO-H$_2$O(b1U"01) | -2726.39441 | -2726.35811 | -2726.35043 | -2726.39120 | -3.08 | -1.59 | -1.62 | 3.90 |
| *$HBrO_2$(MP2)* | *-2723.09431* | *-2723.07844* | *-2723.07358* | *-2723.10547* | - | - | - | - |
| $HBrO_2$ | -2725.07136 | -2725.05584 | -2725.05093 | -2725.08295 | - | - | - | - |
| *$HBrO_2$-$H_2O$(MP2)* | *-2799.37374* | *-2799.33266* | *-2799.32519* | *-2799.36350* | **-11.62** | **-9.19** | **-9.92** | **-0.51** |
| $HBrO_2$-$H_2O$(b'1U11) | -2801.53171 | -2801.49118 | -2801.48366 | -2801.52211 | **-9.86** | **-7.48** | **-8.22** | **1.19** |
| *$HBrO_2$-$H_2O$(MP2)* | *-2799.36545* | *-2799.32596* | *-2799.31748* | *-2799.35823* | *-6.42* | *-4.99* | *-5.08* | *2.81* |
| $HBrO_2$-$H_2O$(b'1U01) | -2801.52198 | -2801.48315 | -2801.47439 | -2801.51624 | -3.75 | -2.44 | -2.40 | 4.87 |
| *$HBrO_2$-$H_2O$(MP2)* | *-2799.36561* | *-2799.32557* | *-2799.31729* | *-2799.35880* | *-6.52* | *-4.74* | *-4.97* | *2.45* |
| $HBrO_2$-$H_2O$(b'1U10) | -2801.52392 | -2801.48463 | -2801.47610 | -2801.51893 | -4.97 | -3.37 | -3.48 | 3.19 |
| *$HBrO_2$-$H_2O$(MP2)* | *-2799.36616* | *-2799.32662* | *-2799.31813* | *-2799.35913* | *-6.87* | *-5.40* | *-5.49* | *2.24* |
| $HBrO_2$-$H_2O$(b"1U10) | -2801.52430 | -2801.48501 | -2801.47658 | -2801.51739 | -5.21 | -3.61 | -3.78 | 4.15 |
| *$HBrO_2$-$H_2O$(MP2)* | *-2799.37352* | *-2799.33240* | *-2799.32494* | *-2799.36323* | *-11.49* | *-9.03* | *-9.77* | *-0.33* |
| $HBrO_2$-$H_2O$(b"1U11) | -2801.53157 | -2801.49102 | -2801.48349 | -2801.52197 | -9.77 | -7.38 | -8.11 | 1.28 |
| | | | | | | | | |
| *$HBrO_3$(MP2)* | *-2798.08840* | *-2798.06924* | *-2798.06332* | *-2798.09796* | - | - | - | - |
| $HBrO_3$ | -2800.22489 | -2800.20639 | -2800.20037 | -2800.23525 | - | - | - | - |
| *$HBrO_3$-$H_2O$(MP2)* | *-2874.36201* | *-2874.31917* | *-2874.30960* | *-2874.35323* | *-7.96* | *-6.50* | *-6.58* | *1.23* |
| $HBrO_3$-$H_2O$(b1U20) | -2876.67890 | -2876.63674 | -2876.62708 | -2876.67091 | -5.88 | -4.35 | -4.44 | 3.39 |
| *$HBrO_3$-$H_2O$(MP2)* | *-2874.36759* | *-2874.32295* | *-2874.31459* | *-2874.35556* | **-11.47** | **-8.87** | **-9.71** | **-0.23** |
| $HBrO_3$-$H_2O$(b1U11) | -2876.68595 | -2876.64220 | -2876.63377 | -2876.67480 | **-10.30** | **-7.77** | **-8.64** | **0.95** |
| *$HBrO_3$-$H_2O$(MP2)* | *-2874.36200* | *-2874.31879* | *-2874.30949* | *-2874.35202* | *-7.96* | *-6.26* | *-6.51* | *1.99* |
| $HBrO_3$-$H_2O$(b'1U'11) | -2876.67767 | -2876.63551 | -2876.62582 | -2876.67024 | -5.11 | -3.58 | -3.65 | 3.81 |
| | | | | | | | | |
| *$HBrO_4$(MP2)* | *-2873.04557* | *-2873.02238* | *-2873.01646* | *-2873.05142* | - | - | - | - |
| *$HBrO_4$(MP2)lower* | *-2873.04583* | *-2873.02242* | *-2873.01577* | *-2873.05240* | *-0.1624** | - | - | - |
| $HBrO_4$ | -2875.34053 | -2875.31847 | -2875.31148 | -2875.34926 | - | - | - | - |
| $HBrO_4$(lower) | -2875.34054 | -2875.31851 | -2875.31150 | -2875.34930 | -0.0034* | - | - | - |
| *$HBrO_4$-$H_2O$(MP2)* | *-2949.32667* | *-2949.27821* | *-2949.26890* | *-2949.31227* | *-12.50* | *-10.17* | *-10.88* | *-1.66* |
| $HBrO_4$-$H_2O$(1U'b11) | -2951.80283 | -2951.75592 | -2951.74638 | -2951.79054 | -11.08 | -8.79 | -9.57 | **-0.11** |
| *$HBrO_4$-$H_2O$(MP2)* | *-2949.32744* | *-2949.27875* | *-2949.26951* | *-2949.31262* | **-12.99** | **-10.51** | **-11.26** | **-1.88** |
| $HBrO_4$-$H_2O$(1Ub11) | -2951.80313 | -2951.75610 | -2951.74656 | -2951.79045 | **-11.27** | **-8.90** | **-9.68** | -0.05 |
| *$HBrO_4$-$H_2O$(MP2)* | - | - | - | - | - | - | - | - |
| $HBrO_4$-$H_2O$(1Ub10) | - | - | - | - | - | - | - | - |
| | | | | | | | | |
| *HIO(MP2)* | *-185.20818* | *-185.19959* | *-185.19564* | *-185.22489* | - | - | - | - |
| *HIO(MP2)lower* | *-185.30469* | *-185.29229* | *-185.28832* | *-185.31729* | *-60.56** | - | - | - |
| HIO | -187.17718 | -187.16945 | -187.16543 | -187.19482 | - | - | - | - |
| HIO(lower) | -187.26885 | -187.25647 | -187.25248 | -187.28147 | -57.52* | - | - | - |
| *HIO-$H_2O$(MP2)* | *-261.57721* | *-261.54127* | *-261.53355* | *-261.57224* | **-7.29** | **-5.90** | *-5.92* | *1.43* |
| HIO-$H_2O$(i1U00) | -263.72435 | -263.68827 | -263.68068 | -263.71901 | **-6.82** | **-5.27** | **-5.37** | 2.21 |
| *HIO-$H_2O$(MP2)* | *-261.57697* | *-261.54015* | *-261.53279* | *-261.57289* | *-7.14* | *-5.20* | *-5.44* | **1.03** |
| HIO-$H_2O$(i1U01) | -263.72299 | -263.68624 | -263.67893 | -263.71850 | -5.96 | -3.99 | -4.27 | 2.53 |
| *HIO-$H_2O$(MP2)* | *-261.57682* | *-261.54014* | *-261.53358* | *-261.57119* | *-7.04* | *-5.19* | **-5.93** | *2.09* |
| HIO-$H_2O$(i1U'01) | -263.72284 | -263.68630 | -263.67885 | -263.71903 | -5.86 | -4.03 | -4.23 | **2.20** |
| *HIO-$H_2O$(MP2)* | *-261.57525* | *-261.53853* | *-261.53127* | *-261.57061* | *-6.06* | *-4.18* | *-4.49* | *2.46* |
| HIO-$H_2O$(i1U"01) | -263.72052 | -263.68408 | -263.67670 | -263.71635 | -4.41 | -2.64 | -2.87 | 3.88 |
| | | | | | | | | |
| *$HIO_2$(MP2)* | *-260.28542* | *-260.26985* | *-260.26492* | *-260.29780* | - | - | - | - |
| $HIO_2$ | -262.42021 | -262.40494 | -262.39996 | -262.43293 | - | - | - | - |
| *$HIO_2$-$H_2O$(MP2)* | *-336.56710* | *-336.52635* | *-336.51883* | *-336.55797* | **-13.04** | **-10.62** | **-11.37** | **-1.84** |
| $HIO_2$-$H_2O$(i'1U11) | -338.88078 | -338.84062 | -338.83293 | -338.87253 | **-10.00** | **-7.70** | **-8.37** | **0.92** |
| *$HIO_2$-$H_2O$(MP2)* | *-336.55992* | *-336.52040* | *-336.51209* | *-336.55298* | *-8.53* | *-6.88* | *-7.13* | *1.29* |
| $HIO_2$-$H_2O$(i'1U01) | -338.87410 | -338.83499 | -338.82655 | -338.86794 | -5.80 | -4.17 | -4.36 | 3.79 |
| *$HIO_2$-$H_2O$(MP2)* | *-336.56063* | *-336.52069* | *-336.51251* | *-336.55431* | *-8.97* | *-7.07* | *-7.40* | *0.45* |
| $HIO_2$-$H_2O$(i'1U10) | -338.87577 | -338.83651 | -338.82809 | -338.87080 | -6.85 | -5.12 | -5.33 | 2.00 |
| *$HIO_2$-$H_2O$(MP2)* | *-336.56240* | *-336.52253* | *-336.51445* | *-336.55480* | *-10.09* | *-8.22* | *-8.62* | *0.15* |
| $HIO_2$-$H_2O$(i"1U10) | -338.87822 | -338.83856 | -338.83054 | -338.87070 | -8.39 | -6.40 | -6.87 | 2.07 |
| *$HIO_2$-$H_2O$(MP2)* | *-336.56686* | *-336.52608* | *-336.51857* | *-336.55768* | *-12.88* | *-10.45* | *-11.20* | *-1.66* |
| $HIO_2$-$H_2O$(i"1U11) | -338.88068 | -338.84052 | -338.83280 | -338.87246 | -9.93 | -7.64 | -8.29 | 0.96 |
| | | | | | | | | |
| *$HIO_3$(MP2)* | *-335.30904* | *-335.29070* | *-335.28445* | *-335.32090* | - | - | - | - |
| $HIO_3$ | -337.61367 | -337.59580 | -337.58945 | -337.62628 | - | - | - | - |
| *$HIO_3$-$H_2O$(MP2)* | *-411.58734* | *-411.54460* | *-411.53516* | *-411.57956* | *-10.91* | *-8.99* | *-9.36* | *-0.90* |
| $HIO_3$-$H_2O$(i1U20) | -414.07321 | -414.03101 | -414.02145 | -414.06603 | -9.35 | -7.40 | -7.76 | 0.83 |
| *$HIO_3$-$H_2O$(MP2)* | *-411.59080* | *-411.54665* | *-411.53827* | *-411.57979* | ***-13.08*** | ***-10.28*** | ***-11.31*** | ***-1.04*** |
| $HIO_3$-$H_2O$(i1U11) | -414.07642 | -414.03295 | -414.02450 | -414.06608 | **-11.37** | **-8.62** | **-9.67** | **0.79** |
| *$HIO_3$-$H_2O$(MP2)* | *-411.58629* | *-411.54311* | *-411.53402* | *-411.57647* | *-10.25* | *-8.05* | *-8.64* | *1.04* |
| $HIO_3$-$H_2O$(i'1U'11) | -414.07064 | -414.02801 | -414.01877 | -414.06168 | -7.74 | -5.52 | -6.08 | 3.55 |
| | | | | | | | | |
| *$HIO_4$(MP2)* | *-410.27638* | *-410.25449* | *-410.24721* | *-410.28727* | - | - | - | - |
| *$HIO_4$(MP2)lower* | *-410.27643* | *-410.25448* | *-410.24729* | *-410.28583* | *-0.03** | - | - | - |
| $HIO_4$ | -412.75468 | -412.73355 | -412.72706 | -412.76382 | - | - | - | - |
| $HIO_4$(lower) | -412.75486 | -412.73359 | -412.72631 | -412.76497 | -0.11* | - | - | - |
| *$HIO_4$-$H_2O$(MP2)* | *-486.55877* | *-486.51160* | *-486.50197* | *-486.54731* | ***-13.449*** | *-11.013* | ***-11.850*** | *-2.663* |
| $HIO_4$-$H_2O$(1U'i11) | -489.21846 | -489.17216 | -489.16243 | -489.20714 | -11.894 | -9.507 | -10.349 | -0.696 |
| *$HIO_4$-$H_2O$(MP2)* | *-486.55877* | *-486.51161* | *-486.50197* | *-486.54742* | ***-13.449*** | ***-11.017*** | ***-11.850*** | ***-2.733*** |
| $HIO_4$-$H_2O$(1Ui11) | -489.21846 | -489.17216 | -489.16244 | -489.20718 | **-11.895** | **-9.513** | **-10.351** | **-0.717** |
| *$HIO_4$-$H_2O$(MP2)* | - | - | - | - | - | - | - | - |
| $HIO_4$-$H_2O$(1Ui10) | - | - | - | - | - | - | - | - |

* – the energy difference in kcal/mol between the isomers.

**Table 2.** The frequencies [$cm^{-1}$] for the supermolecular isomers of water-coupled Sulfurous acid $H_2SO_3$, using B3LYP-DFT/Aug-cc-*p*VDZ level of theory. In the parentheses are given the optimized structure labels given in Table 1 for the $H_2SO_3$ supermolecular isomers.

| $H_2SO_3$ (1U'11sp) | $H_2SO_3$ (1U02sp) | $H_2SO_3$ (1U12sp) | $H_2SO_3$ (1U01sp) | $H_2SO_3$ (1U00sp) | $H_2SO_3$ (1U11sp) |
|---|---|---|---|---|---|
| 49 | 29.(5) | 37 | 17 | 18 | 47 |
| 163 | 29.(9) | 155 | 21 | 27 | 172 |
| 194 | 92 | 196 | 98 | 39 | 197 |
| 343 | 277 | 399 | 319 | 154 | 371 |
| 409 | 334 | 429 | 361 | 314 | 415 |
| 476 | 392 | 474 | 405 | 379 | 477 |
| 795 | 713 | 738 | 728 | 727 | 794 |
| 1046 | 1074 | 1077 | 1075 | 1077 | 1050 |
| 1133 | 1098 | 1157 | 1095 | 1092 | 1136 |
| 3639 | 3700 | 3582 | 3700 | 3698 | 3622 |
| 3707 | 3785 | 3627 | 3749 | 3788 | 3701 |
| 3868 | 3885 | 3871 | 3882 | 3896 | 3860 |

**Figure 1**. The system configurations for optimized geometries (B3LYP/aug-cc-*p*VDZ). The label under each (super)molecular structure corresponds to that in the first column of Table 1.

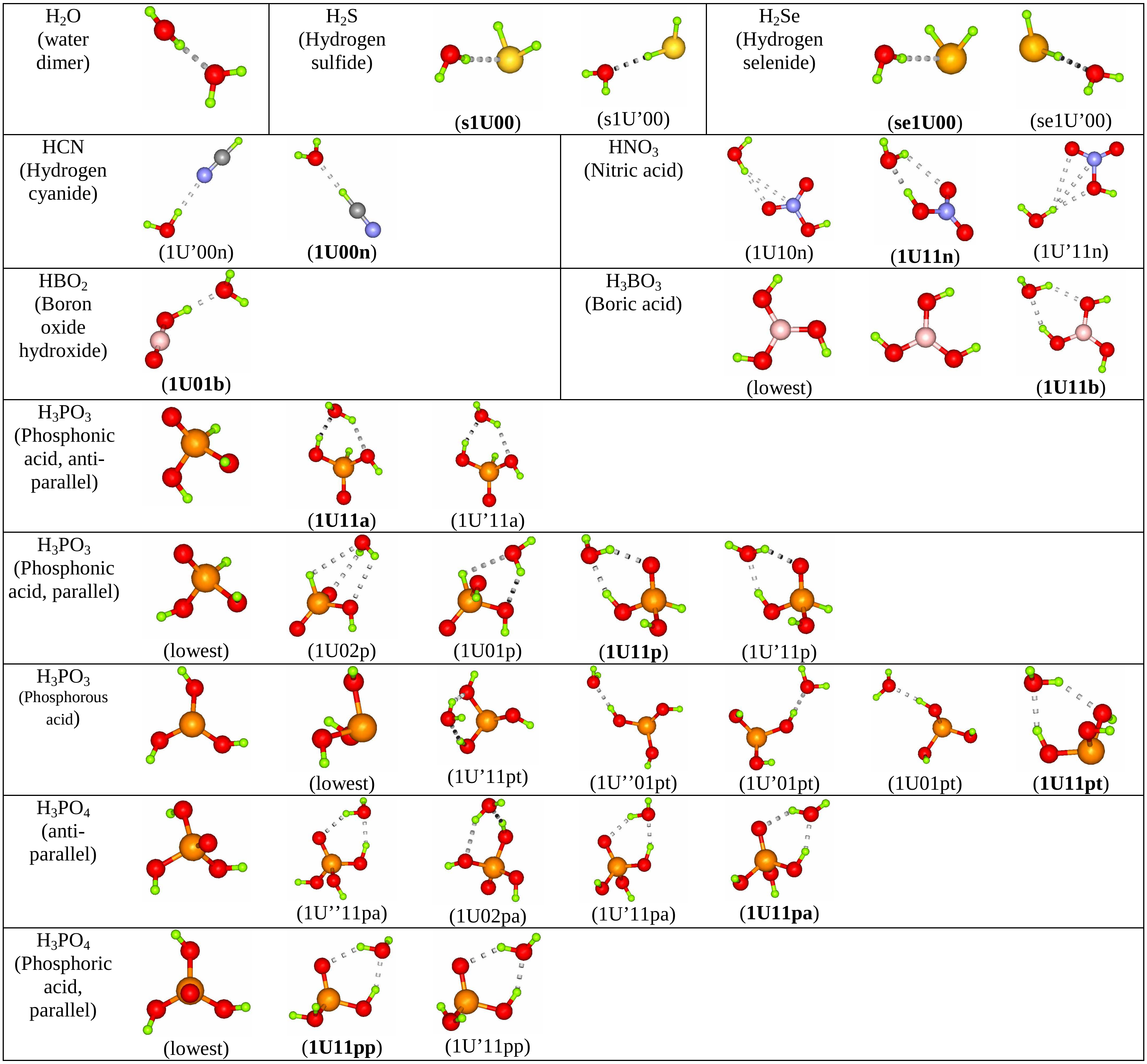

**Figure 1**. (continue) The system configurations for optimized geometries (B3LYP/aug-cc-*p*VDZ). The label under each (super)molecular structure corresponds to that in the first column of Table 1.

| | | | | | | | |
|---|---|---|---|---|---|---|---|
| $H_2SO_3$ (Sulfonic acid) | 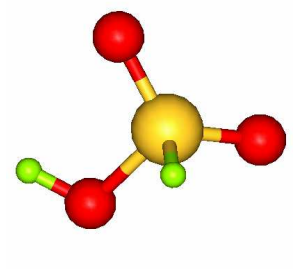 | 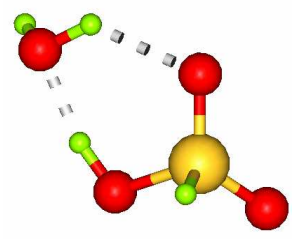<br>(**1U11su**) | 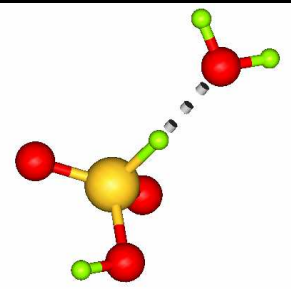<br>(1U00su) | | | | |
| $H_2SO_3$ (anti-parallel) | 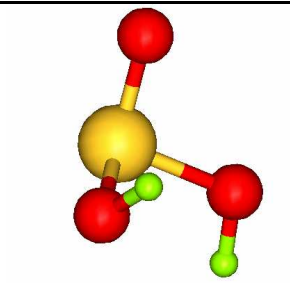 | 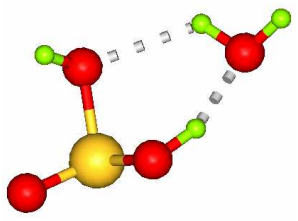<br>(1U02sa) | 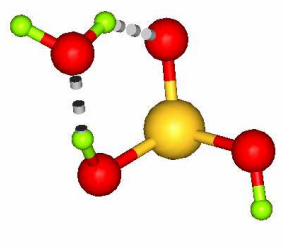<br>(1U’11sa) | 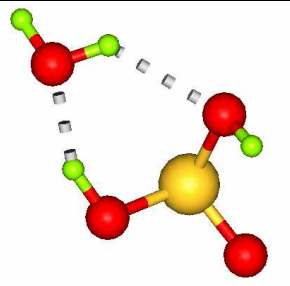<br>(1U’02sa) | 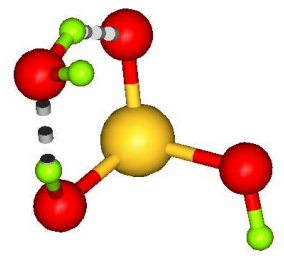<br>(**1U11sa**) | | |
| $H_2SO_3$ (Sulfurous acid, parallel) | 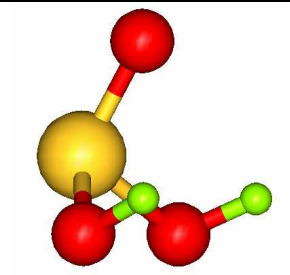<br>(lowest) | 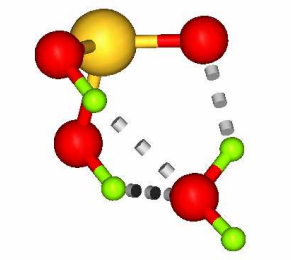<br>(1U12sp) | 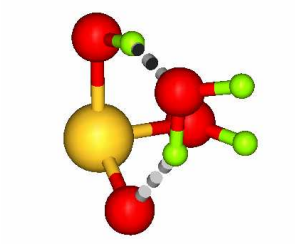<br>(1U’11sp) | 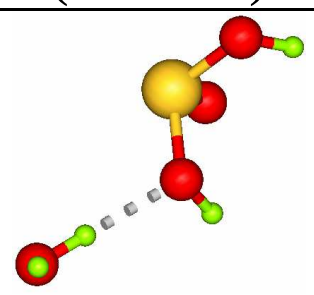<br>(1U01sp) | 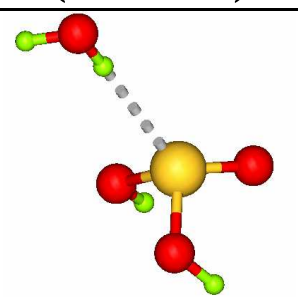<br>(1U00sp) | 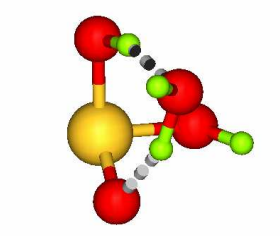<br>(**1U11sp**) | 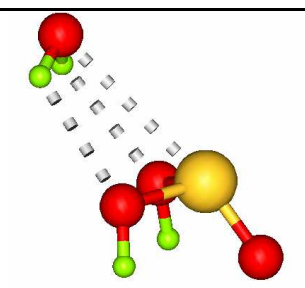<br>(1U02sp) |
| $H_2SO_4$ (Sulfuric acid, anti-parallel, trans) | 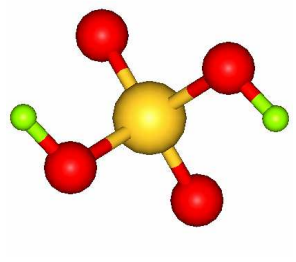<br>(lowest) | 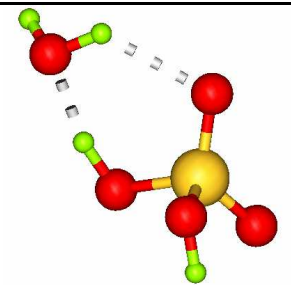<br>(**1U11t**) | 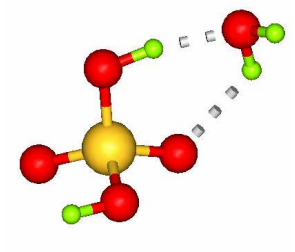<br>(1U’11t) | | | | |
| $H_2SO_4$ (parallel, cis) | 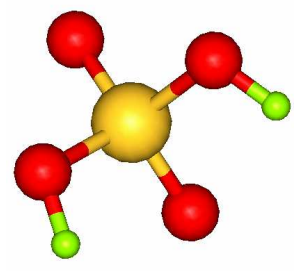 | 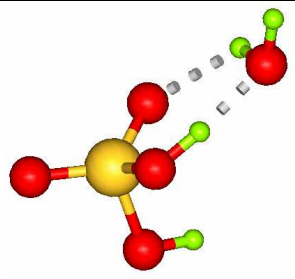<br>(**1U11c**) | 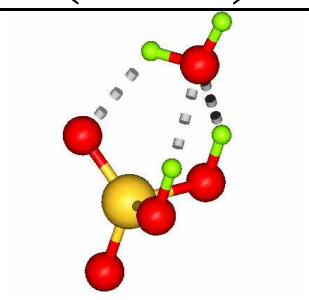<br>(1U12c) | 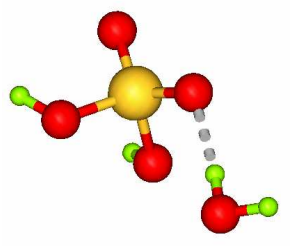<br>(1U10c) | | | |

**Figure2**. The system configurations for optimized geometries (B3LYP/aug-cc-*p*VDZ). The CRENBL ECP effective core potential was used for the heavy iodine atom. The label under each (super)molecular structure corresponds to that in the first column of Table 1.

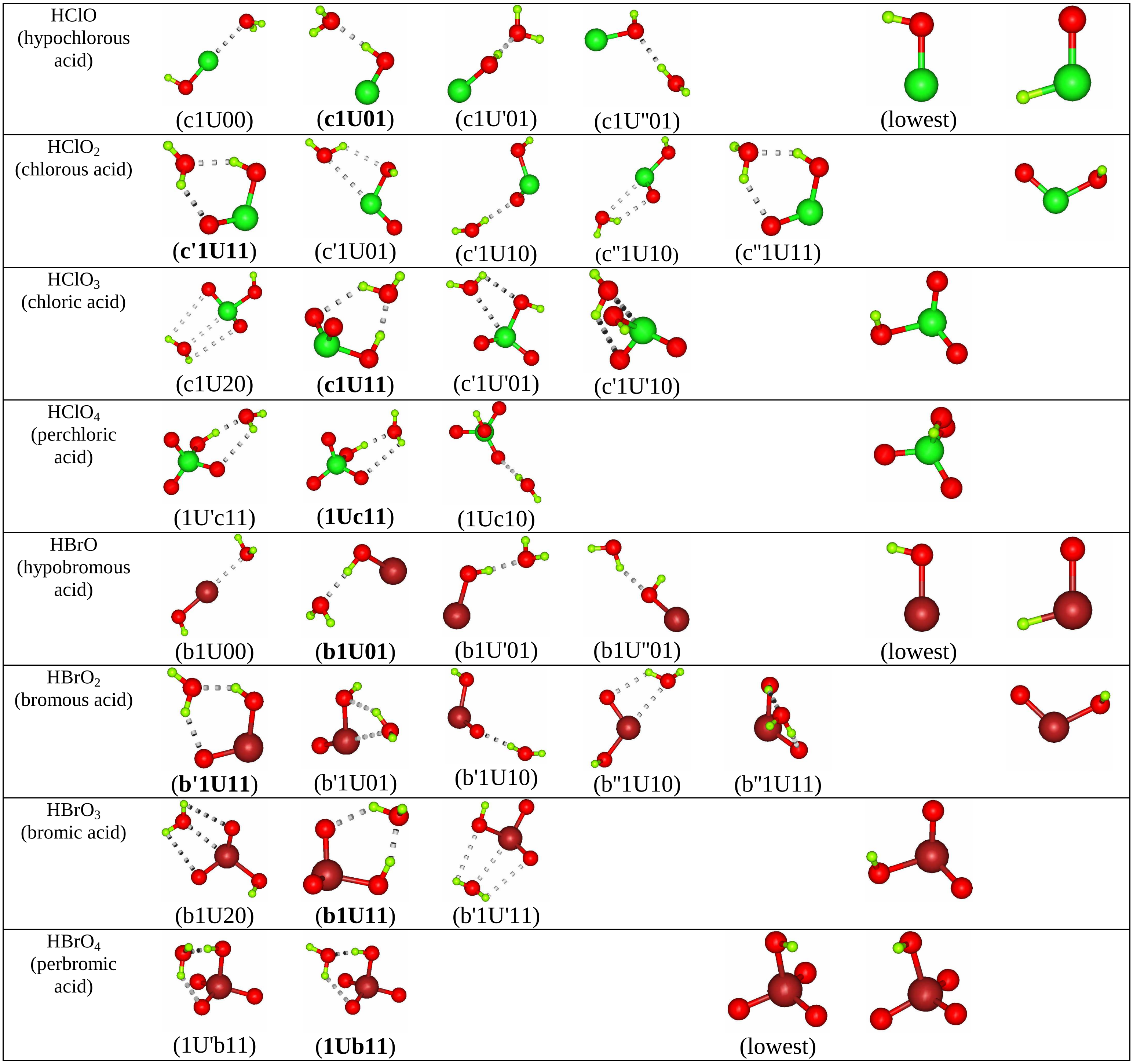

**Figure2**. (continue) The system configurations for optimized geometries (B3LYP/aug-cc-*p*VDZ). The CRENBL ECP effective core potential was used for the heavy iodine atom. The label under each (super)molecular structure corresponds to that in the first column of Table 1.

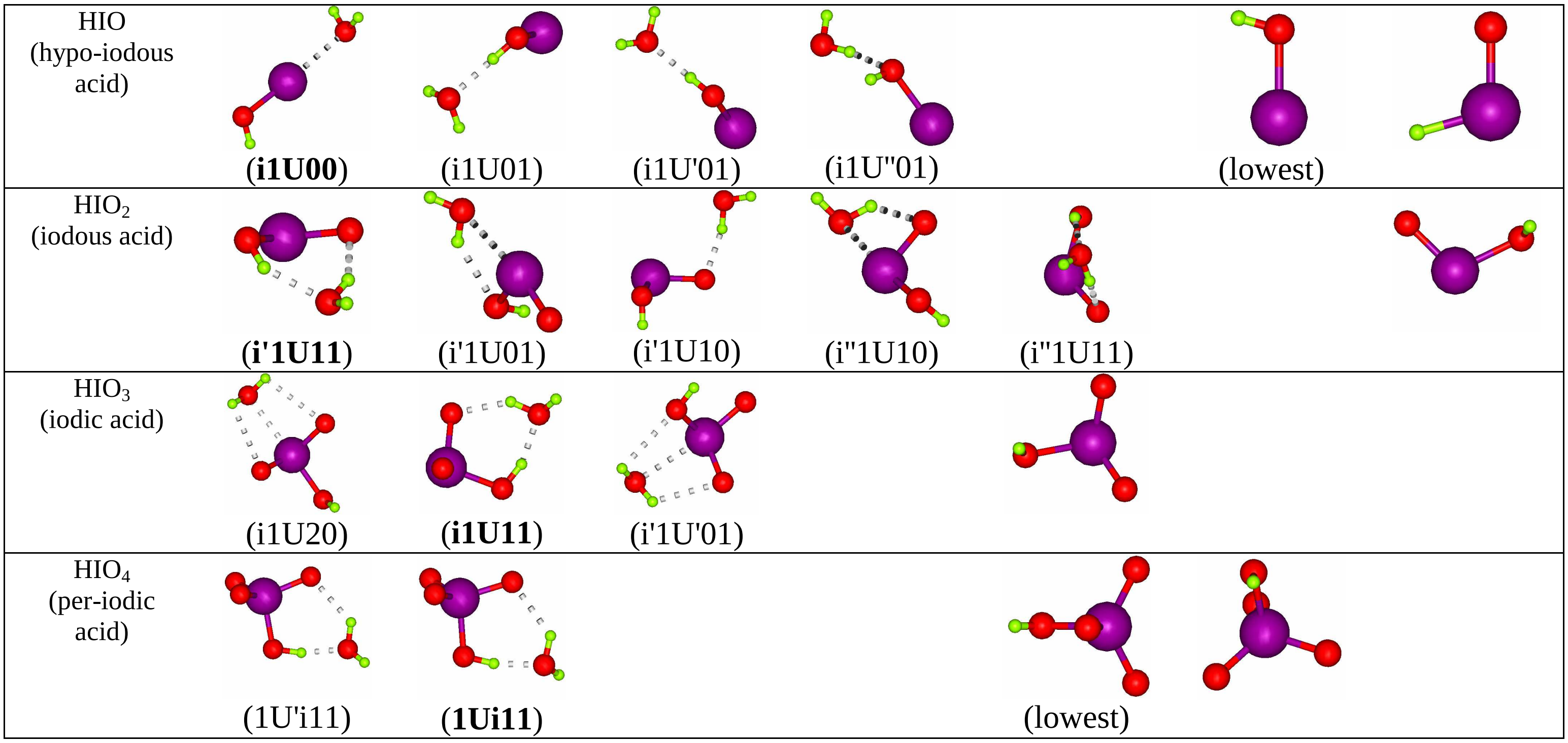